\documentclass[conference]{IEEEtran}
\IEEEoverridecommandlockouts

\usepackage{cite}
\usepackage{amsmath,amssymb,amsfonts}
\usepackage{graphicx}
\usepackage{textcomp}
\usepackage{xcolor}
\def\BibTeX{{\rm B\kern-.05em{\sc i\kern-.025em b}\kern-.08em
    T\kern-.1667em\lower.7ex\hbox{E}\kern-.125emX}}

\usepackage{soul}
\usepackage{enumitem}
\usepackage{flushend}
\usepackage{balance}
\usepackage{algorithm}
\usepackage{algpseudocode}
\usepackage{hyperref}

\usepackage{caption}
\usepackage{float} 
\usepackage{subcaption}
\usepackage{tcolorbox}
\usepackage{diagbox}
\usepackage{multirow}
\usepackage{lipsum}
\usepackage{makecell}
\usepackage{pifont}
\usepackage{engord}
\usepackage{microtype}
\usepackage{amsmath}
\usepackage{booktabs}
\usepackage[table]{xcolor}
\usepackage{bm}
\usepackage{xcolor}
\usepackage{wrapfig}
\usepackage{graphicx} 
\usepackage{placeins}
\usepackage{dblfloatfix}
\usepackage{hhline}

\usepackage{algorithm}
\usepackage{algpseudocode}
\usepackage{amsmath}
\usepackage{multirow, multicol, hhline, colortbl, xcolor}

\newcommand{\tool}{{FlowFixer}}

\begin{document}

\title {Diagnosis-Driven Automatic Repair for Agentic Workflow via Symbolic Inference}





\author{Xuyan~Ma, Yawen~Wang$^{*}$, Junjie~Wang$^{*}$, Xiaofei~Xie,
Boyu~Wu, Mingyang~Li, Dandan~Wang, Qing~Wang$^{*}$

\IEEEcompsocitemizethanks{

\IEEEcompsocthanksitem
X. Ma, Y. Wang, J. Wang, B. Wu, M. Li, D. Wang, and Q. Wang are with State Key Laboratory of Complex System Modeling and Simulation Technology, Institute of Software, Chinese Academy of Sciences, and University of Chinese Academy of Sciences, Beijing, China.
\protect\\
E-mail: \{maxuyan2021, yawen2018, junjie, mingyang2017, dandan, wq\}@iscas.ac.cn, boyu\_wu2021@163.com
\IEEEcompsocthanksitem
X. Xie is with Singapore Management University, Singapore.
\protect\\
E-mail: xfxie@smu.edu.sg
\IEEEcompsocthanksitem
*Corresponding Authors.

}}


\maketitle

\begin{abstract}
Platform-orchestrated agentic workflows have become a popular paradigm for developing LLM-based applications. However, their reliability remains a major challenge due to the uncertainty of LLM outputs, complex inter-node dependencies, and heterogeneous tool interactions. 
Existing agentic workflow optimization and agent enhancement methods primarily rely on trajectory-level feedback. Without explicitly identifying the underlying failure root causes, their resulting repair plans are often insufficiently targeted.
We propose {\tool}, a diagnosis-driven automated repair framework for agentic workflows. {\tool} first transforms workflow executions into unified symbolic traces and performs symbolic inference to derive executable behavioral specifications that capture node correctness, temporal dependencies, and causal relationships. Based on specification verification, it conducts failure attribution and root cause analysis, and then generates targeted repair patches. 
To reduce verification costs, {\tool} further employs a multi-dimensional pre-execution assessment to filter infeasible repairs before dynamic verification. 
We evaluate {\tool} on workflow failures collected from 
three popular development platforms: Dify, Coze and n8n. Results show that {\tool} achieves a repair success rate of 71.3\%, outperforming state-of-the-art baselines by 11.9\% to 27.6\%. It also improves failure attribution accuracy by 4.8\% to 33.1\% and root cause analysis accuracy by 15.3\% to 38.8\%. 
This work offers a new perspective on reliable diagnosis and repair of agentic workflows through symbolic modeling and inference. 
\end{abstract}



\section{introduction}
\label{sec:intro}

Large Language Model (LLM)-based agents have rapidly evolved beyond prompt-based conversational applications into autonomous systems capable of planning, tool use, and multi-step task execution\cite{chowa2026language}. 
Modern development platforms such as Dify \cite{dify}, Coze \cite{coze}, and n8n \cite{n8n} have further popularized agentic workflows, a node-based construction paradigm in which heterogeneous components are connected through explicit control and data dependencies to form executable pipelines. Users can develop workflows in low-code environments by dragging and configuring nodes.
Such platform-orchestrated agentic workflows have been widely adopted in practical LLM applications \cite{leatherwood2025workshop,workflows_dify_2025}, including autonomous task planning\cite{wang2023plan}, software engineering assistants\cite{qian2024chatdev}, and enterprise automation systems\cite{hong2024metagpt}.


Despite their growing adoption, the reliability of agentic workflows remains a concern\cite{hughes2025ai}. 
Unlike traditional software systems, whose execution is largely governed by explicit control flow and deterministic program logic, 
agentic workflows exhibit highly dynamic and uncertain behaviors. 
Such uncertainty arises from probabilistic LLM outputs\cite{ross2025textual}, context-sensitive tool interactions\cite{lymperopoulos2025tools}, and implicit coupling between node configurations, intermediate variables, and downstream execution logic\cite{react}.
As a result, failures are difficult to predict, localize, and repair, since an error in one node may silently propagate through the workflow and manifest as a downstream symptom. 
These failures substantially hinder the reliability, maintainability, and deployment readiness of real-world agentic {workflows}.

{Program repair has long been a central concern in software engineering\cite{lyu2025automatic}. Traditional automated program repair techniques typically leverage issue descriptions, execution traces, and test results to localize defective code regions, generate candidate patches, and validate them against test suites\cite{sweagent,agentless,RepairAgent}. 
However, due to the low-code nature of agentic workflows, such code-centric repair techniques cannot be directly applied to them; furthermore, failures in these workflows may stem from prompts, node configurations, inter-node dependencies, or tool interfaces, rather than from isolated source code issues \cite{bouzenia2025understanding,whowhen}.
Therefore, recent studies have begun to explore repair and optimization techniques for agent systems, including prompt evolution \cite{pei2025scope,hao2026recreate}, workflow restructuring\cite{song2025aegis,wang2025maestro}, trajectory reduction\cite{agentdiet}, and runtime supervision \cite{SUPERVISOR}. 
Most existing techniques formulate agent improvement as an optimization problem, using trajectories, feedback, or final task-level metrics to search for improved system variants. 
Although such optimizations may result in better overall performance, they are often weakly linked to the root causes of concrete failures, leading to coarse-grained and less targeted modifications.
These limitations motivate the need for diagnosis-driven agent repair, where failure attribution and root cause analysis provide explicit guidance for targeted workflow corrections.}

{Realizing such diagnosis-driven workflow repair, however, faces two key challenges.}
{First, workflow failures are only partially observable from execution trajectories.} 
Agentic workflow executions involve probabilistic LLM reasoning, external tool invocations, and complex interactions among heterogeneous nodes\cite{wu2023autogen,shen2023hugginggpt}. As failures propagate through dependent workflow nodes, downstream symptoms are often intertwined with upstream causes, making it difficult to accurately determine where a failure originates and how it evolves throughout the execution process\cite{zhu2025llm,xie2026spark}. 
Second, workflow repair requires reasoning beyond local modifications \cite{xinjie2025reagent}. 
A workflow node is often coupled with prompts, variables, model configurations, and downstream execution behaviors\cite{wu2024agentkit}. 
As a result, a seemingly correct local change may introduce unintended side effects or violate implicit assumptions elsewhere in the workflow.
Effective repair therefore requires not only correcting the localized failure, but also preserves the consistency of the overall workflow logic and task objectives.
{To bridge this gap, we propose \textbf{{\tool}}, a diagnosis-driven repair approach for agentic workflows. 
The core idea is to combine symbolic modeling and symbolic inference to make workflow failures analyzable and repairable.
Specifically, {\tool} first transforms the execution trajectories into unified symbolic representations, making node configurations, node behaviors, inter-node dependencies, and execution states explicitly observable. 
It then performs symbolic inference over the symbolic trace to derive and verify node-level behavioral assertions, producing structured diagnosis evidence for failure attribution and root cause analysis.
{Based on the resulting diagnosis information, {\tool} generates targeted repair candidates that not only address the localized errors, but also takes into account the dependencies of other related nodes as well as the overall task objectives.
Before applying them, it performs a lightweight pre-execution assessment against symbolic constraints of the overall workflow, checking whether each candidate is compatible with inferred node behaviors, preserves inter-node dependencies and execution-state constraints, and remains aligned with the global task objective.
By filtering out infeasible candidates before execution, this assessment reduces unnecessary dynamic verification overhead and leaves only viable candidates for dynamic verification of actual repair effectiveness.}
Furthermore, {\tool} also maintains an experience pool that records online repair feedback and continuously accumulates repair experience, enabling progressively improved failure diagnosis and repair effectiveness across repeated workflow failures.

Experimental results demonstrate the effectiveness of {\tool}. {\tool} achieves a repair success rate of 71.3\%, outperforming all repair baselines by 11.9\% to 27.6\%. In failure diagnosis, it improves failure attribution accuracy by 4.8\% to 33.1\% and root cause analysis accuracy by 15.3\% to 38.8\% over baseline methods. Ablation studies further confirm the effectiveness of symbolic modeling, root cause taxonomy, repair knowledge, execution information, and experience accumulation in improving both diagnosis and repair performance.
In summary, this paper makes the following contributions:
\begin{itemize}
    \item We present a symbolic modeling \& inference based framework for failure diagnosis and automated repair of agentic workflows.
    \item {We propose a unified node behavioral specification that is instantiated as executable assertions to produce structured evidence for failure diagnosis.}
    \item {We construct a multi-dimension assessment mechanism which can effectively eliminate unreasonable modified workflows and reduce verification costs.}
    \item We demonstrate the effectiveness of {\tool} on real-world agentic workflows, showing its capability on accurately failure diagnosis and successfully workflow repair. 
\end{itemize}
\section{Background}
\label{sec:bg}


Modern agentic AI development platforms (e.g., Dify, Coze, n8n) achieve agentic workflow construction through a node-based paradigm. 
Instead of implementing complex logic through source code, users construct workflows by visually connecting predefined nodes and configuring their parameters through graphical user interfaces. A workflow typically consists of multiple interconnected nodes, where each node encapsulates a specific functionality, such as LLM invocation, text processing, tool execution, knowledge retrieval, conditional routing, or data transformation.

The {node types} supported by the platform are summarized as follows:
1) Start and Termination Nodes;
2) LLM and Agent Nodes;
3) Knowledge Nodes;
4) Logic and Control Nodes;
5) Code and Template Nodes;
6) Tool and Integration Nodes.
These nodes extend the platform’s capability enabling workflows to interact with external services, databases or tools. 
Formally, a workflow is defined as:
\begin{equation}
    W = (N, E)
\end{equation}
where $N = \{n_1, n_2,...,n_m\}$ is the set of nodes, $E \subseteq N \times N $ denotes control dependencies between nodes, determining the execution order and information flow. 

Each node $n_i$ is defined as:
\begin{equation}
    n_i = (I_i, T_i,C_i)
\end{equation}
where $I_i$ represents the node name, $T_i$ is node type and $C_i$ is its configuration. The configuration determines the runtime action of the node and varies across different node categories. For example, LLM nodes contain prompts and model parameters; retrieval nodes specify knowledge sources and retrieval strategies; tool nodes define tool selections and invocation parameters; and code nodes contain executable scripts. During workflow execution, nodes transform incoming information ($V^{in}$) into output ($V^{out}$) based on their configurations and propagate the output to downstream nodes:
$V^{out} = C_i(V^{in})$.

\section{Approach}
\label{sec:methodology}
\begin{figure*}[htbp]
    \centering
    \includegraphics[width=\textwidth]{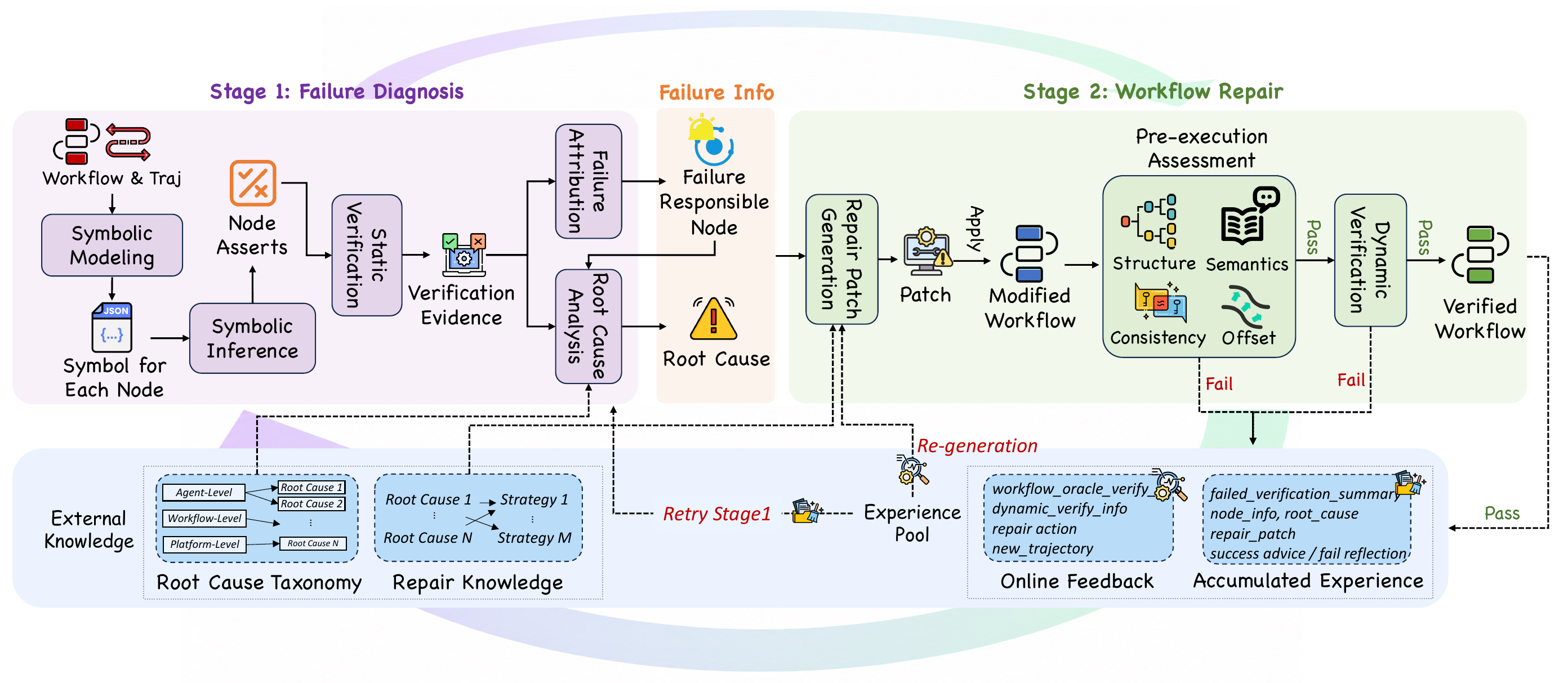}
    \caption{Overview of {\tool}. }
    \label{fig:overview}
    \end{figure*} 
\subsection{Overview}
We propose {\tool}, an agentic workflow automated repair framework leveraging symbolic modeling and inference.
As illustrated in Fig. \ref{fig:overview}, {\tool} contains two core stages: \textbf{Failure Diagnosis}, which identifies the failure-responsible node (i.e., the root node that leads to the final failure) and the root cause, and \textbf{Workflow Repair}, which generates and verifies repair patches to produce repaired workflows. 
{The two stages are tightly coupled through a ``diagnosis $ \leftrightarrow$ repair'' loop.}     


{Specifically, {\tool} first constructs formal behavioral specifications for each workflow node through symbolic modeling and inference.
These specifications are statically checked against the observed execution results to identify violations and collect node-level verification evidence. Based on such evidence and workflow structure, {\tool} locates the
failure-responsible node and analyzes its root cause. 
It then generates repair patches through a set of atomic edit
operations and applies them to the original workflow to obtain a
modified workflow.
The modified workflow undergoes a lightweight pre-execution
assessment that reuses the constructed symbolic specifications to
efficiently filter infeasible patches before costly dynamic execution. 
Repair experiences collected during this process are stored and leveraged in subsequent iterations, enabling a continuous diagnosis and repair refinement loop.}
\subsection{Symbolic Modeling \& Inference}
\label{subsec:symbol}

\subsubsection{\textbf{Symbolic Modeling}}
{When faced with a failed execution trajectory,} in order to eliminate the impact of different trajectory formats on different platform and remove redundant information, we define a unified trajectory format based on symbolic representation.
{\tool} normalizes theses trajectories into a sequence of nodes, $\{node_1, ..., node_n\}$, each node consists of id, type, input \& output, status, configuration contexts and parameters, as shown in the ``Input'' part in Fig. \ref{fig:spec}. 
All nodes are assembled in execution order to form a standardized trace with a unified symbolic format.


\subsubsection{\textbf{Symbolic Inference}}
\label{subsubsec:oracle}

For each node in the trace, {\tool} infers its formal behavioral specifications that characterize the expected behaviors of the node and interactions with the rest of the workflow. 
These specifications are derived from three complementary dimensions that reflect the fundamental correctness requirements of these agentic workflows: \texttt{existence, temporal} and \texttt{causal} constraints.
These dimensions are chosen because workflow failures typically manifest as missing or malformed workflow entities, invalid data/control dependencies, incorrect execution ordering, or causal propagation of upstream errors to downstream symptoms.

Specifically, \texttt{existence constraints} check whether necessary fields, inputs, outputs and configuration elements are present. 
\texttt{Temporal constraints} ensure that node executions follow semantically meaningful orders required by the task, such as retrieving information before summarization or validating inputs before tool invocation. 
\texttt{Causal constraints} characterize the expected semantic relationships among workflow entities, including variable consistency, interface compatibility, task-specific requirements, and causal dependencies between upstream and downstream behaviors. For example, generated travel schedules should satisfy budget limits and duration requirements specified by the task.

This three-dimensional decomposition enables {\tool} to infer from local node validity to workflow-level failure propagation. 
It also provides structured diagnosis signals for subsequent failure attribution and repair, enabling more systematic identification and localization of underlying failure sources. 


\begin{table}[h]
\caption{BNF syntax of DSL.}
\label{fig:dsl}
\centering
\renewcommand{\arraystretch}{1.3}
\begin{tabular}{p{0.23\textwidth} p{0.2\textwidth}}
\toprule
1: $\Phi$ ::= \textbf{assert} Pred & \textit{Top-level assertion} \\
2: Pred ::= Pred \textbf{and} Pred & Boolean logic and comparisons \\
3: \quad $|$ Pred \textbf{or} Pred & \\
4: \quad $|$ \textbf{not} Pred & \\
5: \quad $|$ (Pred) & \\
6: \quad $|$ Expr comp Expr & \\
7: Expr ::= value & Values, variables, field/method \\
8: \quad $|$ var & access \\
9: \quad $|$ var.attr & \\
10: \quad $|$ var.method(args) & \\
11: comp ::= $==, \text{!=}, >, \geqq, <, \leqq$, in, not in & Comparison operators \\
12: edit ::= insert(), remove(), replace(), append(), swap() & Edit operators for patch generation\\
13: args ::= Expr (`,' Expr)* & Argument list (comma-separated) \\
14: var ::= identifier & Valid variable name \\
15: attr ::= id $|$ input $|$ output $|$ $\cdots$ & Of Node object \\
\bottomrule
\end{tabular}
\end{table}








{\tool} uses an LLM-based specification generator to synthesize node-wise behavioral specifications.
Given each node's context, inputs and outputs, workflow connections and the overall task goal, the generator translates the expected node behavior into formal constraints along the three dimensions described above.

A lightweight assertion-based DSL \cite{teoh2026webtestpilot,rabern2026logicskills}, summarized in Table \ref{fig:dsl}, is adopted as the format for inferred specifications. 
It expresses each constraint as a top-level assertion \texttt{assert Pred}, where predicates are constructed from Boolean connectives, comparison operators, and field or method access to workflow entities. 
For example, existence constraints can check required fields through expressions such as \texttt{node.output != None}; 
temporal constraints can compare execution order using order comparisons such as \texttt{<} or \texttt{>}; and causal constraints can combine multiple predicates with \texttt{if}, \texttt{or}, and \texttt{not} to to help determine whether logical relationships are satisfied between nodes.


\begin{figure}[]
    \centering
    \includegraphics[width=\linewidth]{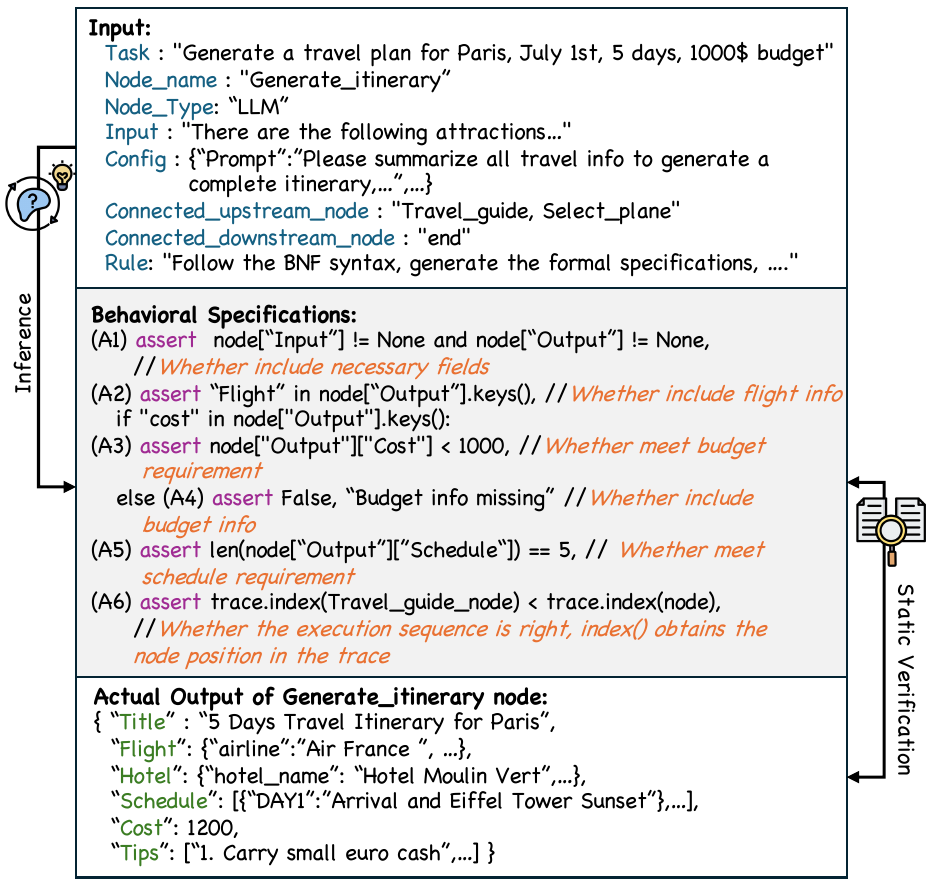}
    \caption{Symbolic Inference Example. ``node'' represents the Generate\_itinerary node for short.
    } 
    \label{fig:spec}
\end{figure}

Fig. \ref{fig:spec} presents an example of the specification inference process. 
For the \textit{Generate\_itinerary} node, the inferred specifications involve several constraints: 
Existence constraints require the generated itinerary cannot be empty (A1);
temporal constraints require \textit{Generate\_ itinerary} node must execute after the upstream node (\textit{Travel\_guide}) has completed its operations (A6); 
and causal constraints ensure consistency with user-provided inputs such as the budget and schedule (A3, A4, A5) while providing necessary itinerary information such as flight details (A2).

\begin{figure*}[htbp]
    \centering
    \includegraphics[width=\textwidth]{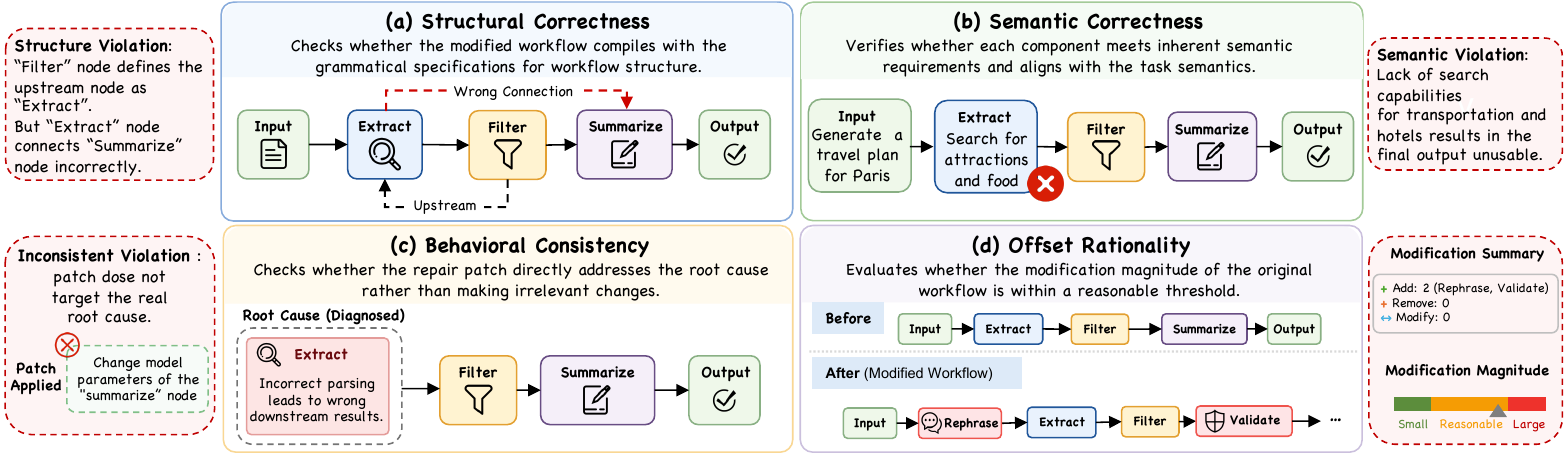}
    \caption{Multi-dimension Pre-execution Assessment.} 
    \label{fig:multi}
    \end{figure*} 

\subsection{Stage 1: Failure Diagnosis}
\label{subsec:failure_analysis}

This stage aims to localize {failure-responsible node} and identify the underlying root causes of a failed trajectory. 
\subsubsection{\textbf{Failure Attribution}}
\label{subsubsec:ranking}




Given the formal behavioral specifications obtained from symbolic inference, {\tool} compares them against the actual outputs of each node. 
By evaluating whether the observed node behaviors satisfy the inferred specifications, it can identify violated behavioral requirements.

For each node, we calculate a suspicious score indicating its contribution to the final failure. The score combines two sources of factors: (1) behavioral abnormality revealed by assertion violations and (2) the node’s potential influence on downstream executions. Specifically, nodes with higher assertion violation rates are more likely to exhibit anomalous behavior and are thus assigned higher suspicious scores. However, assertion violations alone are insufficient because failures may propagate through workflow dependencies and cause downstream nodes to appear abnormal. To account for this effect, we additionally consider the structural position of a node as its potential propagation impact. Nodes located earlier in the workflow can influence a larger portion of subsequent executions and are therefore more likely to be the origin of failures. By combining assertion violations with propagation characteristics, {\tool} prioritizes nodes that are both highly abnormal and capable of explaining downstream failure symptoms.

\subsubsection{\textbf{Root Cause Analysis}}
\label{subsubsec:root_cause_analysis}
The module maps node failures to corresponding root causes (e.g., poor prompt design) and generates structured failure data covering root cause and failure responsible node.

To make root cause analysis more targeted and systematic, {\tool} constructs a root cause taxonomy from prior studies on agent and workflow failures \cite{ma2026demystifying,cemri2026multi,zhu2025llm,lu2025exploring}, as summarized in Fig. \ref{fig:taxonomy}. 
It contains sixteen root cause types with their detailed description grouped into node capability, node orchestration, and node execution, 
and serves as structured prior knowledge for interpreting failure symptoms.
Given the failure-responsible node identified by failure attribution, {\tool} feeds its symbolic context, inferred behavioral specifications, and assertion verification results (include both satisfied and violated assertions) into the analysis module, to produce the root cause type. 
\begin{figure}[]
    \centering
    \includegraphics[width=0.95\linewidth]{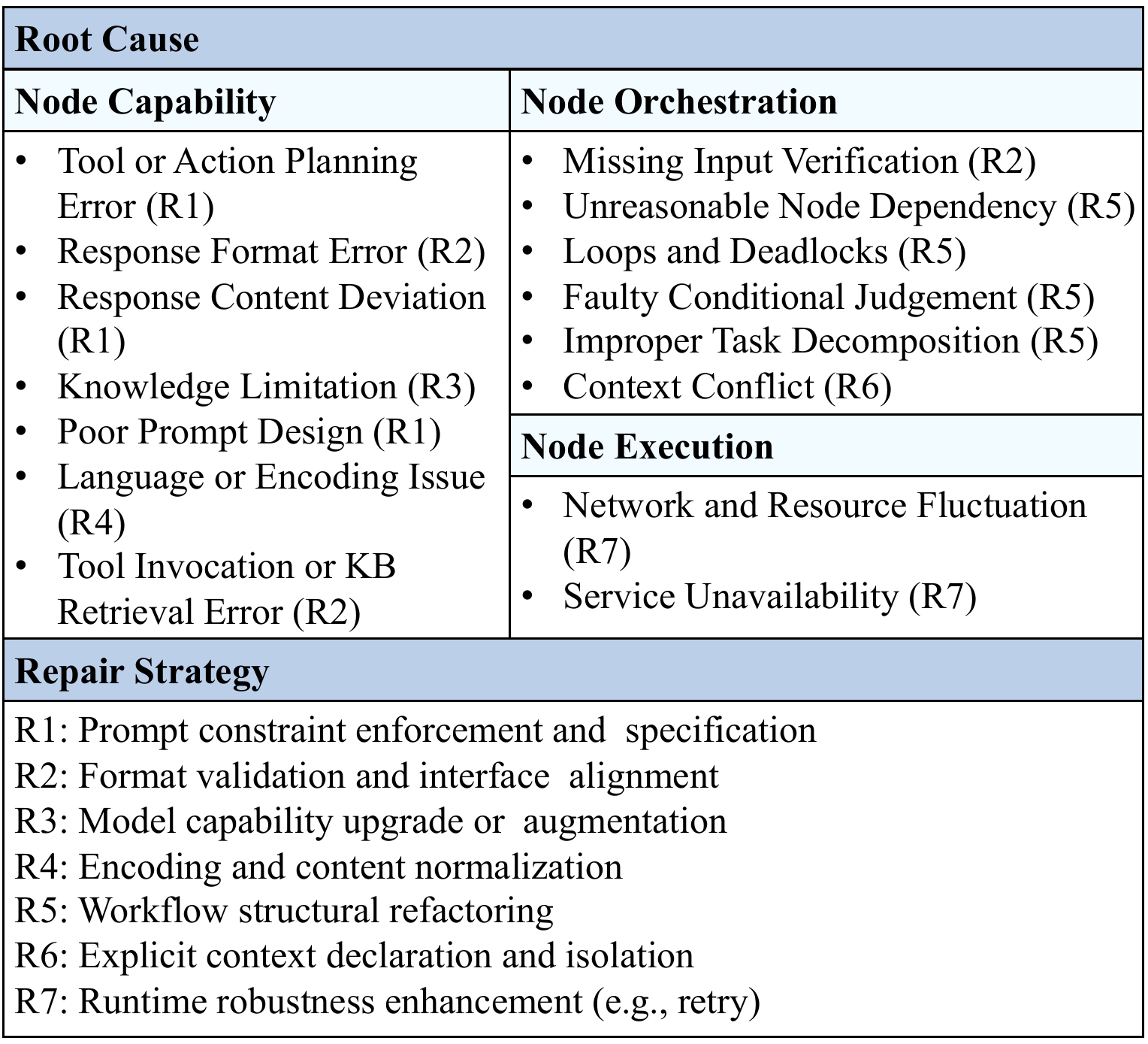}
    \caption{Root Cause Taxonomy and Repair Strategies. Root cause-specific repair strategies are shown in parentheses.} 
    \label{fig:taxonomy}
\end{figure}



\subsection{Stage 2: Workflow Repair}
\label{subsec:workflow_repair}

Given the failure information from Stage 1, the second stage generates, verifies, and applies repair patches to the workflow.

\subsubsection{\textbf{Repair Patch Generation}}
\label{subsubsec:patch_generation}

This module aims to generate repair patches that directly address the failure root cause. 

To make patch generation targeted rather than exploratory,
{\tool} introduces root-cause-aware repair knowledge, which
maps each diagnosed root cause to a corresponding repair strategy, as shown in Fig. \ref{fig:taxonomy}.
These strategies do not directly specify concrete workflow edits. Instead, they guide patch generation by restricting the search space to modifications that are causally relevant to the diagnosed failure. 
Similar to the root cause taxonomy, this repair knowledge is
summarized from prior studies on agent and workflow failures, and platform-specific repair practices \cite{islam2026selfheal,AgentFixer,ma2026demystifying}.

Based on this knowledge, {\tool} first determines the repair
direction for the diagnosed failure.
For instance, poor prompt design is addressed through prompt constraint enforcement, response format errors through format validation and interface alignment. 
Guided by the selected repair strategy, {\tool} performs patch planning using the failure-responsible node, its symbolic context, violated behavioral specifications, and upstream/downstream dependencies. 
To represent concrete modifications, {\tool} defines five atomic edit operators:

\begin{itemize}
    \item \textbf{Insert.} Insert a new node or node configuration into the workflow at a specific position to supplement missing logic or data processing steps.
    \item \textbf{Remove.} Remove redundant, wrong, or unnecessary nodes and node configurations from the workflow to eliminate invalid execution branches.
    \item \textbf{Replace.} Replace an existing faulty node or node configuration with a correct or more appropriate alternative to fix functional or semantic errors.
    \item \textbf{Append.} Add a new node or node configuration to the end of the workflow to extend execution logic or supplement post-processing steps.
    \item \textbf{Swap.} Exchange the execution order of two nodes to correct causal or sequential errors in the workflow.
\end{itemize}

A repair patch is represented as: 
\begin{equation}
    \text{Patch} = \{ \text{edit}_1, \text{edit}_2, \dots, \text{edit}_k \},
\end{equation}

where each edit can be instantiated as: 

\begin{equation}
    \text{edit} = op(target, content, position)
\end{equation}

Here, $op$ is one of the atomic edit operators, $target$ is the workflow component to be modified, $content$ is the newly inserted or replacement content, and $position$ specifies where the edit should be applied. After generation, the patch is applied to the original workflow to produce a modified workflow.

\subsubsection{\textbf{Pre-execution Assessment}}
\label{subsubsec:static_verification_repair}
To minimize verification costs by eliminating problematic workflows before actual execution, {\tool} performs workflow assessment from four different perspectives. 

\textbf{Structural Correctness}  
     ensures that the workflow can be correctly parsed and executed without structural errors such as missing nodes, invalid connections, or malformed node configuration information. 
   The assessment criteria are directly derived from the workflow platform specifications and node definitions. These predefined structural rules ensure that the repaired workflow conforms to valid workflow syntax and execution constraints.
   
    
    \textbf{Semantic Correctness} verifies whether the repaired workflow satisfies both node-level functional requirements and task-level semantic objectives. 
    For this assessment, {\tool} first masks the actual configurations of the modified nodes and provides the LLM with the formalized workflow representation (mentioned in Sec. \ref{subsubsec:oracle}) and the task description, which jointly specify the execution context of the modified nodes and the overall semantic objective. 
    Based on these information, LLM infers the expected configuration specifications for the modified nodes and subsequently compares them against the nodes' actual configurations in the repaired workflow to assess semantic correctness.

    \textbf{Behavioral Consistency}  assesses whether the generated repair actions are aligned with the diagnosed root cause and the behavior specifications. 
    {\tool} first summarizes the repair patch into a high-level description and then provides the diagnosed root cause and the violated behavioral specification to LLM. Based on these information, it reasons whether the proposed modifications are causally relevant to the identified failure causes and are expected to address the violated specifications rather than introducing unrelated changes.

    \textbf{Offset Rationality} evaluates whether the modification magnitude of the original workflow caused by the repair patch is within a reasonable threshold. The modification should neither be excessively large (which may introduce new defects or deviate from the original workflow logic) nor excessively small (which may fail to effectively fix the identified root cause).

\subsubsection{\textbf{Dynamic Verification and Iterative Loop}}
\label{subsubsec:dynamic_verification}

Patches that pass pre-execution assessment proceed to dynamic verification, where the modified workflow is executed against the original test inputs. 
If the workflow executes successfully, the patch is accepted as the final repair.


If the generated repair fails the verification stage, {\tool} does not immediately terminate the repair process. Instead, it will retry to generate new patch or the workflow is fed back into the failure diagnosis stage. The diagnosis-repair cycle continues until either a valid repaired workflow is obtained or the retry budget is exhausted. If the retry count exceeds the test budget, the workflow is regarded as a failed repair. The corresponding diagnosis and repair experiences are still recorded in the experience pool to support future debugging and repair tasks.

\subsection{Experience Pool}
\label{subsec:experience_pool}

To improve long-term performance, {\tool} maintains an \textit{Experience Pool} that stores execution information and  summarized repair knowledge:

\textbf{Online Repair Feedback}: feedback collected during the current repair process, such as pre-execution assessment results, applied repair operations, and execution trajectories of modified workflows.
    
\textbf{Accumulated Repair Experience}: knowledge distilled
from historical cases, such as recurring failure patterns, root-cause associations, effective repair strategies, and lessons from failed attempts.


After each repair cycle, {\tool} stores both online repair feedback and accumulated repair experience in the experience pool. When subsequent failures occur, relevant records are retrieved as auxiliary knowledge. Online repair feedback serves as short-term memory within the current repair process, capturing the outcomes of previous repair attempts and helping the framework avoid repeating ineffective modifications. In contrast, accumulated repair experience serves as long-term knowledge distilled from historical cases, providing reusable insights into recurring failure patterns, root-cause associations, and effective repair strategies. Together, they support both failure diagnosis and repair generation in future repair iterations.

\section{Experiment}
\label{sec:exp}

\subsection{Research Questions}
\label{subsec:rq}

We consider the following research questions:

\textbf{RQ1: (Repair Effectiveness)} Can {\tool} effectively repair failures for agentic workflow? 

\textbf{RQ2: (Diagnosis Effectiveness)} Can {\tool} effectively conduct failure attribution and identify root cause?

\textbf{RQ3: (Ablation Study)} Can each designed component in {\tool} contribute to failure diagnosis and repair?

\subsection{Dataset}
\label{subsec:dataset}
We adopt the publicly available AgentFail\cite{ma2026demystifying} dataset as the primary experimental benchmark for our evaluation, which consists of 307 annotated failure logs collected from 10 real-world agentic systems built on popular workflow orchestration platforms, Dify and Coze. It covers diverse task domains such as information retrieval, task planning, and code generation, and each data record contains complete execution trajectories, original workflow configurations and fine-grained expert annotations of root causes of the failures.

To further increase the diversity of workflow platforms and failure patterns, we additionally collect 136 failure cases from the n8n platform\cite{n8n}, covering various tasks, including information retrieval, task planning and automated design assistant. 
Similarly, each collected case contains the workflow configuration, execution trajectory and manually annotated root causes, following the annotation pipeline in AgentFail. 

In the subsequent result and analysis section, we merge the results from two datasets since the methods have similar results on the datasets.


\subsection{Baselines}
\label{subsec:baseline}

To evaluate the workflow repair performance, we choose six methods as baselines which can be divided into two categories. 

The first category focuses on traditional program repair: 
1) \textbf{SWE-agent} \cite{sweagent}, which is a popular agent resolving issues in software code repositories.
2) \textbf{RepairAgent} \cite{RepairAgent} treats the LLM as an agent capable of autonomously planning and executing actions to fix bugs by invoking tools. 

The second category focuses on bug repair for agents or agent  evolution and enhancement:
3) \textbf{Recreate} \cite{hao2026recreate} is based on the complete interactive experience of agents. It uses reasoning to identify the causes of success or failure and modifies agent's architecture. 
4) \textbf{GEPA} \cite{GEPA} enables LLM to directly read and reflect on the trajectories during the task execution, thereby facilitating more efficient learning. 
5) \textbf{Scope} \cite{pei2025scope} achieves the agent’s prompt evolution by balancing tactical specificity (resolving immediate errors) with strategic generality (evolving long-term principles).
6) \textbf{SelfHeal} \cite{islam2026selfheal} consists of fix agent and critic agent, leveraging fix rules and web search to automatically repair agent bugs.

Furthermore, to evaluate the failure attribution performance of {\tool} (Stage 1), we select representative methods for trajectory failure attribution as baselines for comparison:
1) \textbf{FAMAS} \cite{ge2025introducing} locates the responsible agents through trajectory replay and spectrum analysis. 
2) \textbf{Dover} \cite{ma2512dover} is an intervention-driven debugging framework, which augments hypothesis generation with active verification through targeted interventions. 
3) \textbf{AgentFixer} \cite{AgentFixer} constructs fifteen failure-detection tools and two root-cause analysis modules. 

\subsection{Evaluation Metrics}
\label{subsec:metrics}
To comprehensively evaluate the performance of {\tool} in workflow repair and failure diagnosis, we use the following three metrics:

\textbf{Repair Success Rate (RSR)}: The ratio of cases where the modified workflow passes the dynamic verification.

\textbf{Failure Attribution Accuracy (FAA)}: The ratio of cases where the method correctly locates the specific node to the total number of failure cases. 

\textbf{Root Cause Accuracy (RCA)}: It refers to the proportion of cases in which the root cause is correctly identified among the cases where failure attribution is correctly predicted.



\subsection{Experiment Setup}
\textit{\textbf{The Overall Setup.}}
In all experiments, we use {\tool} and baselines to respectively repair workflows in the experimental dataset. For each case in dataset, repair process is repeated three times to avoid randomness, and the average performance is presented. All experiments are conducted on a server with an Intel Xeon E5-2690 v4 CPU, 128GB RAM, and an NVIDIA A100 GPU. 
For the LLM involved in {\tool}, we adopt GPT-5.2 as the backbone model.




For \textbf{RQ3}, we assess the effectiveness of designed component, including symbolization, external knowledge and the experience pool. 
The detailed settings of variants are as follows:
\begin{itemize}
    \item \textbf{w/o symbol}: This variant performs failure diagnosis and workflow repair based on the original raw trajectory.
    \item \textbf{Variants related to external knowledge}: We evaluate three variants by progressively removing external knowledge sources. \textbf{w/o Taxonomy} removes the predefined root cause taxonomy during diagnosis, \textbf{w/o Repair} removes the repair knowledge used for patch generation, and \textbf{w/o External} removes both components, disabling all external knowledge throughout the diagnosis and repair process.
    \item \textbf{Variants related to the experience pool}: We further evaluate the contribution of the experience pool through three variants. \textbf{w/o Feedback} removes online repair feedback, preventing iterative optimization between diagnosis and repair. \textbf{w/o Experience} removes accumulated repair experience while retaining online repair feedback. \textbf{w/o Pool} removes both components, relying solely on the execution trajectory for failure diagnosis and repair.
\end{itemize}


\section{Results and Analysis}
\label{sec:result}

\subsection{RQ1: Repair Effectiveness of {\tool}}
\label{subsec:RQ1}


Table \ref{tab:RQ1} presents the overall performance comparison between {\tool} and the baselines on workflow repair (column 2). Overall, {\tool} consistently achieves the best performance, demonstrating the effectiveness of symbolic inference-guided workflow repair.


{\tool} achieves a repair success rate of 71.3\%, substantially surpassing SWE-agent (43.7\%) and RepairAgent (51.2\%), yielding absolute improvements of 27.6\% and 20.1\%, respectively. This result indicates that directly applying program repair techniques designed for traditional software systems is insufficient for agentic workflows, whose failures often arise from unpredictable outputs and complex interactions between workflow structures, rather than deterministic code defects.

{\tool} also consistently outperforms agent enhancement methods. Compared with each baseline, FlowFixer improves repair success rates by 11.9\% (ReCreat), 12.5\% (GEPA), 18.4\% (Scope), and 15.9\% (SelfHeal), respectively. Although these methods leverage self-reflection, prompt refinement, or iterative improvement mechanisms, they primarily focus on enhancing overall agent performance rather than explicitly diagnosing workflow failures. Such approaches lack explicit modeling of workflow semantics and causal relationships, limiting their ability to accurately identify root causes and generate targeted repair solutions. Consequently, they often make blind modifications without accurately identifying the specific workflow components responsible for the failures. Moreover, these methods mainly improve agent performance through prompt optimization, while many issues originate from node configurations, or inter-node dependencies, which cannot be effectively repaired through prompt modifications alone.

Overall, these results demonstrate that symbolic modeling \& inference jointly enable {\tool} to generate more precise and effective repairs for complex agentic workflows.
\newcommand{\impr}[1]{\,{\scriptsize(+#1)}}

\begin{table}[tbp]
\centering
\caption{Performance comparison. (RQ1 \& RQ2, values in parentheses indicate the relative improvement of {\tool} over the baselines.)}
\label{tab:RQ1}
\renewcommand\arraystretch{1.4}
\resizebox{0.48\textwidth}{!}{
\begin{tabular}{c|ccc}
\toprule
Method & Repair (RSR) & Failure Attribution (FAA) & Root Cause (RCA) \\
\midrule

\rowcolor{gray!15}
\multicolumn{4}{c}{\textit{Traditional Automatic Program Repair}} \\

\textbf{SWE-agent}
& 43.7\% \impr{27.6\%}
& 51.9\% \impr{32.5\%}
& 49.1\% \impr{38.8\%} \\

\textbf{RepairAgent}
& 51.2\% \impr{20.1\%}
& 64.4\% \impr{20.0\%}
& 61.7\% \impr{26.2\%} \\

\rowcolor{gray!15}
\multicolumn{4}{c}{\textit{Agent Enhancement}} \\

\textbf{ReCreat}
& 59.4\% \impr{11.9\%}
& 70.9\% \impr{13.5\%}
& 69.8\% \impr{18.1\%} \\

\textbf{GEPA}
& 58.8\% \impr{12.5\%}
& 69.6\% \impr{14.8\%}
& 68.3\% \impr{19.6\%} \\

\textbf{Scope}
& 52.9\% \impr{18.4\%}
& 71.3\% \impr{13.1\%}
& 72.6\% \impr{15.3\%} \\

\textbf{SelfHeal}
& 55.4\% \impr{15.9\%}
& 68.0\% \impr{16.4\%}
& 70.1\% \impr{17.8\%} \\

\midrule

\rowcolor{gray!15}
\multicolumn{4}{c}{\textit{Failure Attribution}} \\

\textbf{FAMAS}
& -
& 51.3\% \impr{33.1\%}
& - \\

\textbf{DOVER}
& -
& 79.6\% \impr{4.8\%}
& - \\

\textbf{AgentFixer}
& -
& 66.8\% \impr{17.6\%}
& 62.1\% \impr{25.8\%} \\

\hline\hline

\textbf{{\tool}}
& \textbf{71.3\%}
& \textbf{84.4\%}
& \textbf{87.9\%} \\

\bottomrule
\end{tabular}}
\end{table}




\subsection{RQ2: Diagnosis Effectiveness of {\tool}}
Table \ref{tab:RQ1} also presents the performance about  failure diagnosis. 
Since FAMAS and DOVER only locate nodes that introduce failures without further analyzing underlying root causes, these two methods can only be evaluated in terms of FAA and do not support RCA calculation.

FAMAS achieves only 51.3\% FAA, substantially lower than the 84.4\% achieved by {\tool}. As a spectrum-based fault localization approach, FAMAS identifies suspicious execution steps according to the correlation between failures and execution traces. It lacks an understanding of workflow semantics and cannot infer the functional roles of workflow nodes or the dependencies between them.
Although DoVer leverages intervention-based execution analysis to achieve good failure attribution performance (79.6\%), it relies on repeatedly modifying and replaying agent behaviors to estimate causal effects, incurring substantial computational overhead. Moreover, {\tool} consistently outperforms DoVer, indicating that precise failure attribution can be achieved without resorting to costly intervention-based analysis.
AgentFixer achieves 66.8\% FAA and 62.1\% RCA. This is because, although it diagnoses agent failures from multiple perspectives and offers valuable insights into their root causes, its analysis operates primarily at the trajectory level; it relies on the overall execution trajectory rather than performing fine grained reasoning focused on workflow nodes and their semantic dependencies.






\subsection{RQ3: Ablation Study}
\label{subsec:rq4}

\newcommand{\imp}[1]{\,{\scriptsize(+#1)}}
\begin{table}[tbp]
\centering
\caption{Ablation study of {\tool}.}
\label{tab:ablation}
\renewcommand\arraystretch{1.4}
\resizebox{0.48\textwidth}{!}{
\begin{tabular}{c|ccc}
\toprule
Method & Repair (RSR) & Failure Attribution (FAA) & Root Cause (RCA) \\
\midrule

\textbf{w/o Symbol}
& 47.0\% \impr{24.3\%}
& 59.8\% \impr{24.6\%}
& 58.3\% \impr{29.6\%} \\

\rowcolor{gray!15}
\multicolumn{4}{c}{\textit{External Knowledge Related}} \\

\textbf{w/o Taxonomy}
& 53.5\% \impr{17.8\%}
& 80.2\% \impr{4.2\%}
& 54.4\% \impr{33.5\%} \\

\textbf{w/o Repair}
& 60.2\% \impr{11.1\%}
& 78.9\% \impr{5.5\%}
& 70.4\% \impr{17.5\%} \\

\textbf{w/o Knowledge}
& 51.1\% \impr{20.2\%}
& 75.0\% \impr{9.4\%}
& 52.7\% \impr{35.2\%} \\

\rowcolor{gray!15}
\multicolumn{4}{c}{\textit{Experience Related}} \\

\textbf{w/o Online}
& 54.2\% \impr{17.1\%}
& 77.7\% \impr{6.7\%}
& 72.2\% \impr{15.7\%} \\

\textbf{w/o Experience}
& 48.4\% \impr{22.9\%}
& 62.5\% \impr{21.9\%}
& 70.9\% \impr{17.0\%} \\

\textbf{w/o Pool}
& 41.7\% \impr{29.6\%}
& 55.1\% \impr{29.3\%}
& 61.4\% \impr{26.5\%} \\

\hline\hline

\textbf{{\tool}}
& \textbf{71.3\%}
& \textbf{84.4\%}
& \textbf{87.9\%} \\

\bottomrule
\end{tabular}}
\end{table}

\textbf{Effect of Symbolization}. Removing symbolization causes substantial performance degradation across all three tasks, with repair success rate dropping by 24.3\%, failure attribution accuracy by 24.6\%, and root cause analysis accuracy by 29.6\%. This result demonstrates that symbolic modeling and symbolic inference provide the foundation for accurate failure diagnosis and workflow repair by making workflow semantics and inter-node dependencies explicitly analyzable.

\textbf{Effect of External Knowledge}. We progressively remove the root cause taxonomy, repair knowledge, and both components. The results show that the root cause taxonomy mainly contributes to accurate root cause analysis, whose removal reduces RCA by 33.5\%. In contrast, repair knowledge primarily improves repair generation, and removing it decreases the repair success rate by 11.1\%. Together, they bridge failure diagnosis and workflow repair, enabling {\tool} to translate diagnosed failure causes into effective repair strategies.

\textbf{Effect of Experience Pool}. We further evaluate the contributions of online repair feedback, accumulated repair experience, and the entire experience pool. The results indicate that the two components play complementary roles. Online repair feedback enables iterative refinement between diagnosis and repair, and removing it decreases the repair success rate by 17.1\%. In contrast, accumulated repair experience provides reusable knowledge from historical cases, and removing it reduces failure attribution accuracy by 21.9\%. Together, they enable {\tool} to continuously refine failure diagnosis and generate more effective repair strategies through experience-guided repair.



\section{Case Study} 
\begin{figure*}[htbp]
    \centering
    \includegraphics[width=0.97\textwidth]{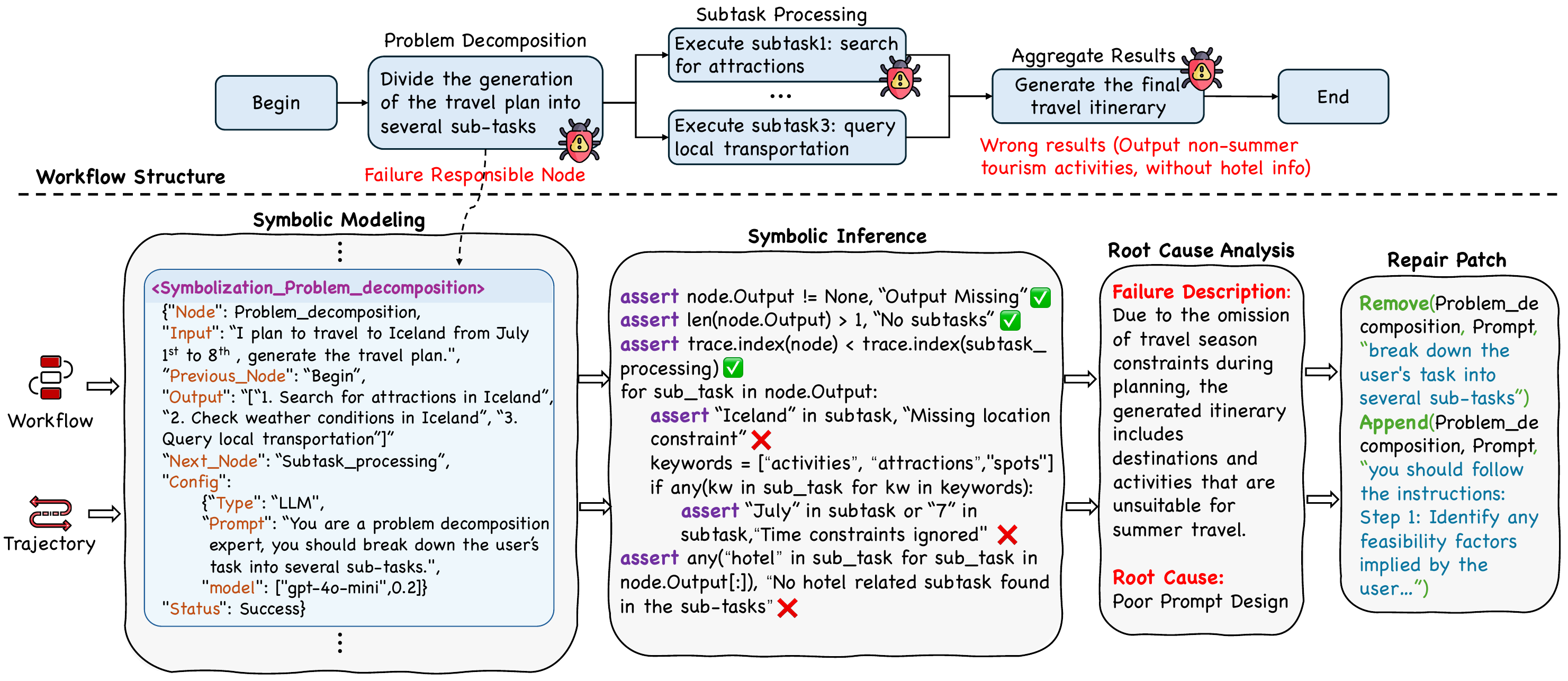}
    \caption{Example of the process of failure diagnosis and repair.
    }
    \label{fig:example}
    \end{figure*} 
{Fig. \ref{fig:example} and \ref{fig:example2} illustrate two examples, in which the first example focuses on the process of symbolic modeling and inference; The second one focuses on demonstrating the process of generating the patch and verifying modified workflow.}

\textbf{Diagnosis Case}. This workflow is for generating a summer travel itinerary for Iceland. However, the final output contains activities that are unsuitable for summer travel and lacks hotel-related information.

In the failure diagnosis, {\tool} performs symbolic modeling and inferring for each node.
Taking the ``Problem Decomposition'' node as an example, {\tool} derives that its formal behavioral specifications should require: the generated sub-tasks should include location information; adhere to user-specified time constraints; and cover essential travel-related elements such as hotel planning.

By verifying these specifications against actual output, {\tool} identifies multiple violations, including missing time constraints (the summer travel period) and the absence of hotel-related sub-tasks. 
Combining these violation evidences with the workflow, {\tool} attributes the failure to the ``Problem Decomposition'' node and identifies the root cause as ``Poor Prompt Design'' as the prompt fails to instruct the model to preserve critical travel constraints during task decomposition.

Guided by the diagnosed root cause and the corresponding repair strategy, {\tool} generates a repair patch consisting of atomic workflow edits. In this example, the repair removes the original decomposition prompt and appends additional instructions requiring the model to explicitly consider user-specified feasibility factors and travel constraints. {\tool} then produces a modified workflow based on the patch and submits it to the assessment and verification stage.

\begin{figure}[]
    \centering
    \includegraphics[width=0.95\linewidth]{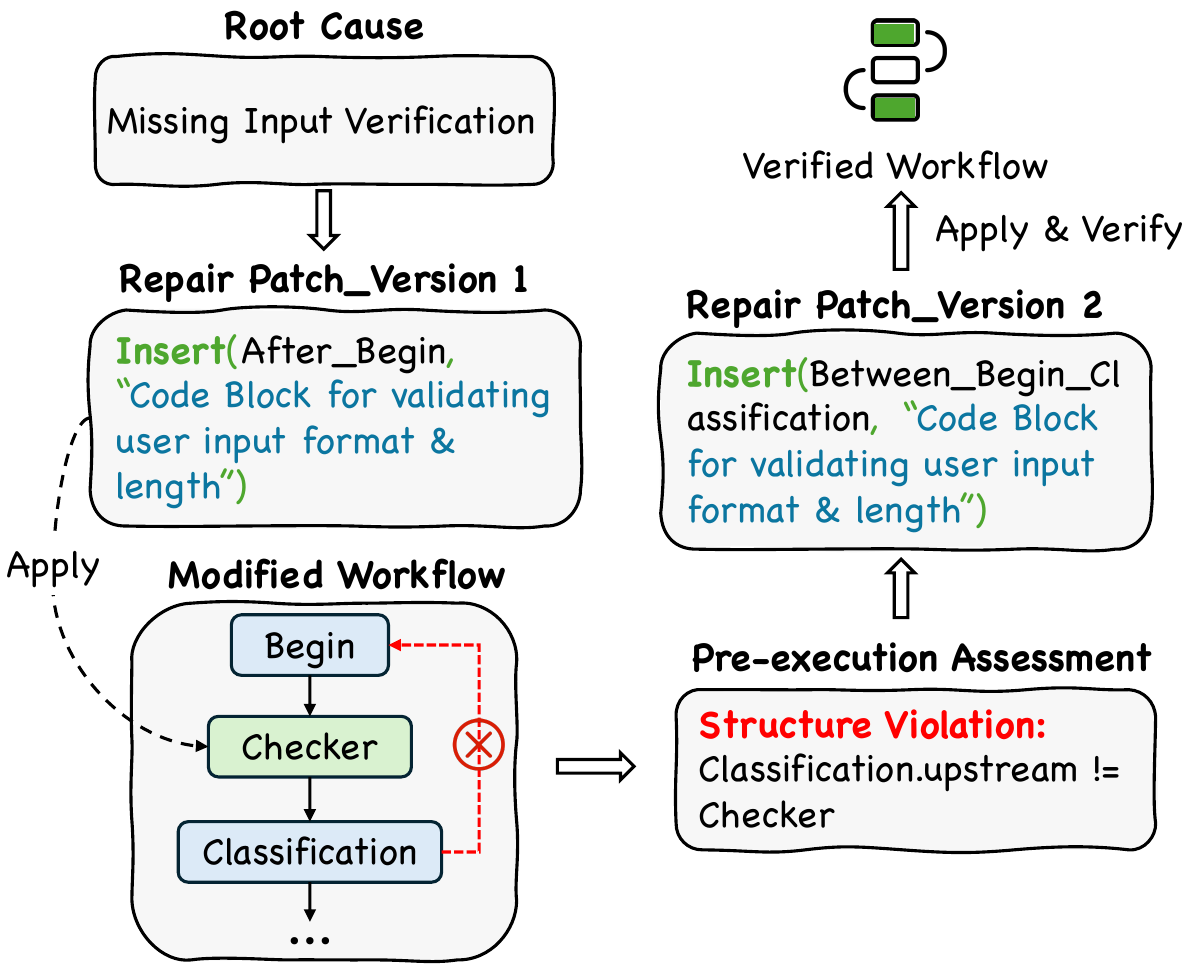}
    \caption{Example of the process of repair.} 
    \label{fig:example2}
\end{figure}

{\textbf{Repair Case}. This workflow implements a Q\&A assistant. However, when users provide inputs with an invalid format or excessively long queries, the workflow lacks an input validation mechanism, causing downstream nodes to receive unexpected inputs and ultimately leading to a crash. }

{Guided by the corresponding repair strategy, {\tool} first generates a repair patch by inserting a validation node immediately after the ``Begin'' node. After applying the patch, the pre-execution assessment analyzes the modified workflow and detects a structure violation: the upstream dependency of the ``Classification'' node has not yet been changed from ``Begin'' to the newly inserted ``Checker'' node, breaking the expected workflow structure. Consequently, this repair candidate is rejected before expensive dynamic execution.}

{{\tool} then generates a new repair patch by inserting ``Checker'' between ``Begin'' and ``Classification'', preserving the original execution dependencies while introducing the required input validation. The repaired workflow successfully passes the pre-execution assessment and is subsequently verified through dynamic execution, producing a valid repaired workflow.}

\section{Discussion}
\label{sec:discuss}

\subsection{Effectiveness of Pre-execution Assessment}
In this section, we evaluate whether the proposed pre-execution assessment can accurately identify wrong repair candidates before costly dynamic execution. We compare its assessment results (\textit{pass}/\textit{fail}) with the corresponding dynamic execution results to measure its ability to filter out invalid workflows while retaining correct repair candidates.
We use \textbf{Precision} and  \textbf{Recall} to evaluate the effectiveness of pre-execution assessment: Precision quantifies the accuracy of pre-execution failure assessment, Recall reflects the detection coverage of real wrong workflows.



Experimental results show that pre-execution assessment module achieves a \textbf{99.7\% Precision}, indicating that almost all workflows identified as wrong by the assessment indeed fail during dynamic execution. This demonstrates that the assessment can effectively filter out low-quality repair candidates with very few false alarms, thereby avoiding unnecessary execution costs caused by obviously incorrect workflows.

The assessment further achieves a \textbf{Recall} of \textbf{84.6\%}, suggesting that the majority of invalid workflows can be detected before execution. Nevertheless, a small portion of failures remain undetected. These cases typically involve model capability limitation or external tool invocation that cannot be fully revealed through static assessment alone and only manifest during actual workflow execution.

These results indicate that {\tool} can effectively eliminate most invalid repair candidates prior to dynamic validation while retaining correct workflows, substantially reducing the cost of workflow repair and verification.


\subsection{Repair Condition Analysis}

We further analyze the repair patches generated by {\tool} from the perspectives of repair operators and modified workflow nodes, as shown in Fig. \ref{fig:discuss}.

\begin{figure}[t]
    \centering
    \includegraphics[width=0.95\linewidth]{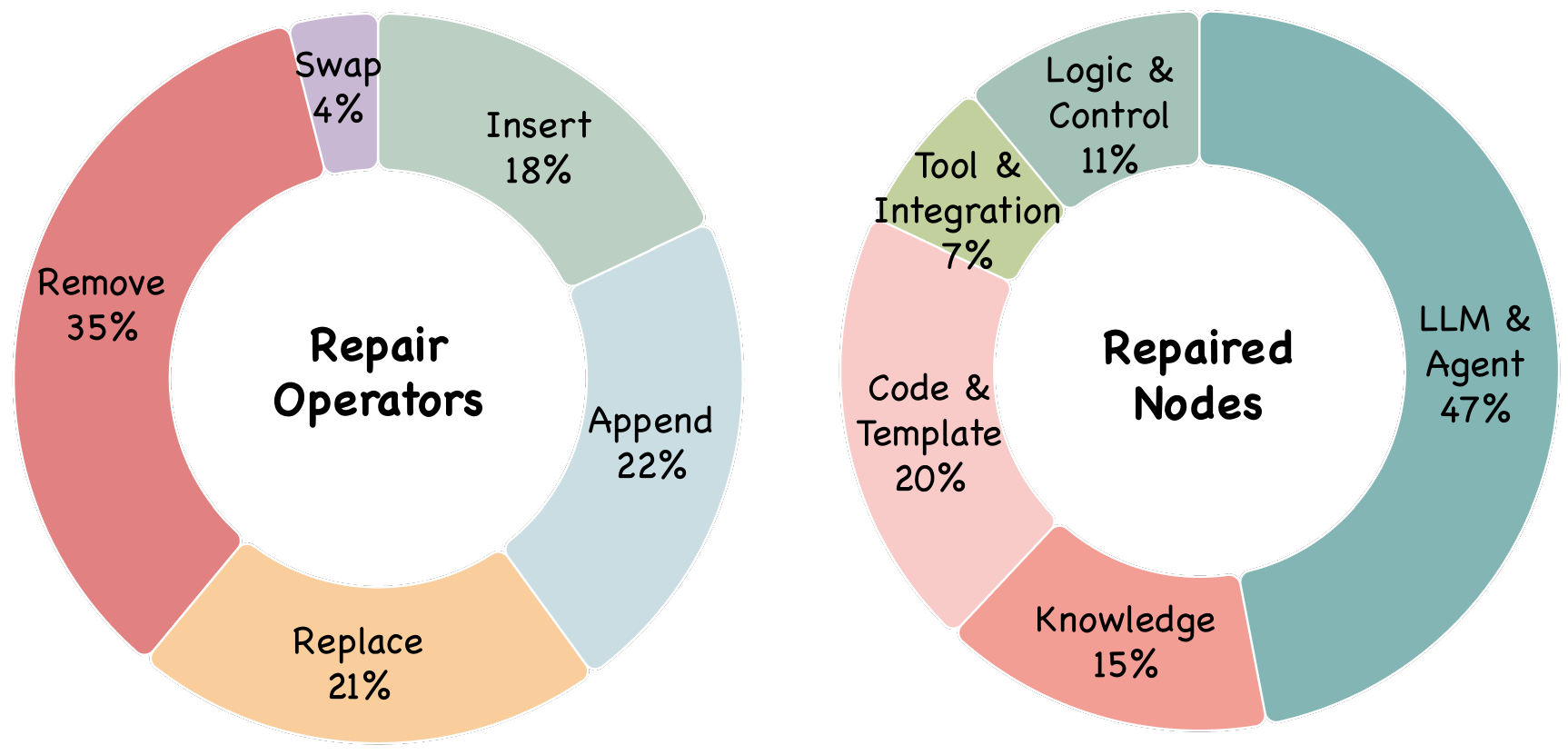}
    \caption{Distribution of repair operators and repaired node.} 
    \label{fig:discuss}
\end{figure}

    

\textbf{Repair Operators}. We can observe that \textit{Remove} (35\%) is the most frequently used operator, \textit{Append} (22\%), \textit{Replace} (21\%), and \textit{Insert} (18\%) are also widely adopted, while even \textit{Swap} (4\%) is required in a subset of repairs. Rather than relying on a single editing strategy, {\tool} employs diverse repair operators to accommodate different failure scenarios. This observation suggests that repairing agentic workflows is inherently complex and requires flexible combinations of workflow modifications, instead of being addressed through a single type of repair operation.

\textbf{Repaired Nodes}. \textit{LLM \& Agent Nodes} account for the largest proportion of repairs (47\%), followed by \textit{Code \& Template Nodes} (20\%), \textit{Knowledge Nodes} (15\%), \textit{Logic \& Control Nodes} (11\%), and \textit{Tool Integration Nodes} (7\%). The repaired nodes are distributed across all major workflow components, indicating that failures can arise from diverse parts of agentic workflows rather than being concentrated in a single node category. This result highlights the necessity of diagnosis and repair techniques that can reason over heterogeneous workflow components and support diverse repair operations.

\subsection{Repair Cost}
We further evaluate the computational cost of {\tool} by measuring the average token consumption throughout the repair process. 
Among the repair methods, Scope incurs the lowest cost at around 14K tokens per repair, while SWE-agent has the highest cost at approximately 38K tokens. In comparison, {\tool} requires 17K tokens on average, indicating a competitive computational overhead.
The relatively low overhead of {\tool} is largely attributed to the proposed pre-execution assessment module, which filters out infeasible repair candidates before workflow execution. By eliminating low-quality modifications in advance, {\tool} avoids unnecessary workflow execution and reduces the overall repair cost. We also observe that the remaining overhead mainly comes from symbolic inference and iterative repair retries. In future work, we plan to try more efficient framework (such as symbolic inference techniques combined with inference caching) to further reduce cost while maintaining repair effectiveness.
\section{Related Work}
\label{sec:rw}
\subsection{Agent Enhancement}
With the widespread application of agents, some new tasks have emerged, aiming to enhance the capabilities of these agents specifically to address the failures they encounter in practical applications.
Existing agent enhancement techniques can be generally categorized into three mainstream paradigms: structural \& workflow enhancement \cite{song2025aegis,wang2025maestro,CEgraph,costa2025instruction}, agent internal enhancement\cite{pei2025scope,zhang2026agentdevel,hao2026recreate} and runtime \& supervisory optimization\cite{agentdiet,SUPERVISOR}.

The first category aims to enhance system robustness by optimizing the external architecture: 
Aegis\cite{song2025aegis} aimed to improve agent-environment observability and computation offloading; 
Maestro \cite{wang2025maestro} optimized workflow graphs and agent configurations guided by trajectory feedback; 
CE-Graph \cite{CEgraph} refined workflows based on failure distribution to avoid repeated errors. 
The second line upgrades the agent’s intrinsic capabilities: 
SCOPE\cite{pei2025scope} achieved online prompt evolution to balance error correction and long-term strategy improvement; 
AgentDevel\cite{zhang2026agentdevel} formalized self-improvement as a release-like pipeline with diagnostic scripting and gating mechanisms; 
ReCreate\cite{hao2026recreate} constructed domain agents by extracting experience from successful and failed trajectories. 
The last line enhances stability during execution: 
AgentDiet\cite{agentdiet} compressed redundant trajectory information to improve efficiency; 
SUPERVISORAGENT\cite{SUPERVISOR} added a lightweight supervisor to proactively correct misbehavior and purify observations. 

\subsection{Failure Attribution}
Failure attribution is the cornerstone of diagnosing and improving LLM-based agent systems. 
Deshpande et al.\cite{deshpande2025trail} introduced a dataset and error taxonomy for evaluating agent traces, showing that current LLMs failed significantly at automated debugging.
Current research on failure attribution in LLM-based agent systems can be broadly categorized into four types: pattern-based analysis\cite{ge2025introducing,barrak,yu2025correct}, LLM reasoning-based methods\cite{whowhen,banerjee2025did,zhu2026raffles}, model fine-tuning techniques\cite{zhang2025agentracer,zhang2025graphtracer,kong2025aegis}, and intervention-based methods\cite{ma2512dover,zhu2025llm,chen2026seeingelephantbenchmarkfailure}.

For pattern-based methods, FAMAS\cite{ge2025introducing} is a spectrum-based failure attribution technique that ranks the suspiciousness of agents based on their occurrence patterns in successful versus failed executions.
For the second category, Who\&When\cite{whowhen} is a representative approach that employs prompting strategies such as all-at-once and step-by-step reasoning to identify the responsible agent and failure step directly from execution logs.
In model fine-tuning-based methods, AgenTracer\cite{zhang2025agentracer} constructs a dedicated failure attribution dataset through fault injection and counterfactual replay, and fine-tunes a lightweight model to recognize error propagation patterns and localize failures efficiently.
As a dynamic intervention-based method,
{DoVer}\cite{ma2512dover} formulates failure attribution as an intervention-driven debugging process, validating attribution hypotheses by modifying intermediate states and observing whether the failure can be eliminated.
\section{Conclusion}

This paper presents {\tool}, a diagnosis-driven automated repair framework for platform-orchestrated agentic workflows. By combining symbolic modeling and symbolic inference, {\tool} enables accurate failure diagnosis and targeted workflow repair. Experimental results on real-world workflow failures show that {\tool} consistently outperforms existing repair and diagnosis approaches in terms of repair success rate, failure attribution accuracy, and root cause analysis accuracy. These results demonstrate the effectiveness of diagnosis-driven symbolic inference for improving the reliability and maintainability of agentic workflows.


\bibliographystyle{IEEEtran}
\bibliography{IEEEfull}

\end{document}